# A note on "Constraints on deep-seated zonal winds inside Jupiter and Saturn"

Gary A. Glatzmaier

glatz@es.ucsc.edu

Earth and Planetary Sciences Department, University of California, Santa Cruz, CA 95064

## Abstract

Liu et al. 2008 discuss an important consideration for models of zonal winds deep within giant planets. However, the constraints they propose for the depth of the winds are based on their prescriptions for the internal structures of the magnetic field and zonal winds. The same kinematic analysis applied to other plausible configurations would produce no constraint on the depth to which the winds extend.

Keywords: (1) MAGNETIC FIELDS, (2) ATMOSPHERES, DYNAMICS, (3) INTERIORS

## 1. Introduction

Liu et al. 2008 propose constraints on the depths to which the zonal winds on Jupiter and Saturn extend below their surfaces. They argue, using analyses based on ohmic heating and Taylor-Proudman constraints, that the observed zonal winds on the surfaces of these planets extend down to no further than 0.96 of Jupiter's radius and 0.86 of Saturn's radius. Constraining the total ohmic heating to not exceed the observed luminosity is certainly important for any interior model of a planet or star. This has been discussed for three-dimensional (3D) magnetohydrodynamic (MHD) models of the geodynamo (e.g. Roberts and Glatzmaier 2000, Christensen and Tilgner 2004) and models of ice giants (e.g. Holme and Bloxham 1996). The issue is more complicated for gas giants for which the density and electrical conductivity vary in radius by many orders of magnitude. The depth-dependence of electrical conductivity in Jupiter and Saturn is nicely described in Liu et al. Their analysis of the amplitude of zonal flows below the surface of a giant planet is based on their chosen internal magnetic field, their chosen pattern for the internal angular velocity and on their simple estimates of the 3D nonlinear vector interactions. This *note* does not argue against their conclusions, but points out some problems with their approach and thus the uncertainty of their conclusions.

## 2. Ohmic heating constraint

Liu et al. 2008 estimate the electric current density in the interior of a giant planet and then integrate its square to estimate the total ohmic heating. They consider only the axisymmetric part of the magnetic field and only the axisymmetric zonal (east-west) flow, i.e., a differential rotation relative to the mean rotation rate of the planet. They estimate the internal poloidal magnetic field by extrapolating the observed axisymmetric field above the surface into the interior, assuming a steady-state internal potential (i.e., curl-free) field. They then prescribe the differential rotation they wish to test and estimate how it shears their internal poloidal field into a toroidal (east-west) field balanced by magnetic diffusion. Integrating in radius then gives them an estimate of the latitudinal component of the electric current density, which together with their estimate of the depth-dependent electrical conductivity allows them to integrate in volume for the total rate of ohmic heating. Simply put, if one makes their assumptions, local ohmic heating is proportional to the square of the local dot product of the poloidal magnetic field and the gradient of the angular velocity. This means that if surfaces of constant angular velocity were everywhere parallel to the internal potential field total ohmic heating would vanish. This is a version of the *Ferraro Iso-rotation Law* (Ferraro 1937).

The main conclusions of Liu et al. 2008 are based on their predicted ohmic heating rates plotted in Fig. 3, which are based on their downward-extrapolated curl-free poloidal magnetic field and their downward-extrapolated "constant on cylinders" angular velocity. That is, they prescribe an internal angular velocity that is only a function of cylindrical radius. This is based on the assumption that the Coriolis and pressure gradient forces are the main forces and that they balance to first order. It is also similar to what current 3D computer simulations of convection in giant planets typically produce (e.g. Christensen 2002).

However, since the approach of Liu et al. 2008 is purely kinematic, one might instead choose an angular velocity constant on the internal poloidal magnetic field lines, which as mentioned above, would induce absolutely no toroidal field, no current density and therefore no ohmic heating (according to the steady-state axisymmetric analysis of Liu et al.). Such a differential rotation profile would have angular velocity nearly "constant on cylinders" at low latitude and nearly "constant on radii" at middle and high latitudes, similar to what is inferred for the sun's interior based on helioseismology (e.g. Thompson et al. 1996) and what is seen in high-resolution 3D simulations of convection in the sun (e.g. Elliott et al. 2000). In Jupiter, a more "constant on radii" profile may be expected in the shallow regions because, besides the Coriolis and pressure gradient forces, the divergence of Reynolds stress is more significant there due to the smaller density scale heights and larger convective velocities.

Liu et al. 2008 consider a variation on this problem in their section 4. Instead of extrapolating the external potential field throughout the deep interior, they prescribe different profiles for the poloidal field above and below a specified depth. Above their specified depth they assume, as before, the poloidal field is a potential field. Therefore, the issue of ohmic heating in this region is the same as that discussed above. That is, because the curl of the extrapolated poloidal field vanishes everywhere above their specified depth, electric current density and ohmic heating are produced only where this field has a component parallel to the *gradient* of the angular velocity.

Their Figs. 6 and 8 would show no ohmic heating in this region for an angular velocity profile aligned with the poloidal magnetic field lines.

Below this specified depth (section 4b in Liu et al. 2008) they assume some unspecified dynamo mechanism is maintaining an axisymmetric magnetic field that is everywhere parallel to the rotation axis with an intensity that varies only in cylindrical radius. In addition, they assume a "constant on cylinders" angular velocity profile for this deeper region. Their Eq. 9 gives zero ohmic heating in this region because the field is aligned with surfaces of constant angular velocity. However, unlike the field above, here the field has a non-zero curl, which provides a longitudinal component of electric current density (their Eq. 16). They use this to get their prediction of ohmic heating as a function of the specified depth for this region (their Eq. 17). However, their results, plotted in their Fig. 7, indicate that this field-flow configuration would generate less ohmic heating than the observed luminosity if the specified depth of this interior region were *below* 0.97 of Jupiter's radius or *below* 0.93 of Saturn's radius. That is, the *deeper* this region is the *less* ohmic heating there would be because, for a given electric current (determined by the curl of the prescribed magnetic field), ohmic heating decreases as electrical conductivity increases. Actually, this part of their example places no constraint on the amplitude of the prescribed "constant on cylinders" angular velocity because the angular velocity contributes nothing to ohmic heating in this deep region.

The example worked out in the appendix of Liu et al. is mathematically interesting but irrelevant to giant planets. The prescribed angular velocity in that example is "constant on spherical surfaces" (their Eq. 44) - not a "cylindrical flow" as stated in their Fig. 12 caption. This is an extreme example of a prescribed magnetic field (an axial dipole) *not* being aligned with the independently prescribed surfaces of constant angular velocity, which therefore maximizes ohmic heating.

**3. Taylor-Proudman constraint**
The second approach Liu et al. 2008 present is an estimate of the amplitudes of the terms in the vorticity equation that might be able to break the Taylor-Proudman tendency to maintain a "constant on cylinders" angular velocity profile (Proudman 1916). The problem with this approach is that a scale analysis, based on simple estimates of the 3D turbulent flow amplitudes and nonlinear correlations below the surface, is too speculative. Instead of solving dynamically consistent equations, Liu et al. use mixinglength arguments based on the assumption of non-axisymmetric convective columns (Ingersoll and Pollard 1982). However, the classic description (Busse 1983) of long thin convective columns spanning continuously through the northern and southern hemispheres and deformed by the sloping boundaries is likely inaccurate for these strongly-turbulent density-stratified gas giants (Glatzmaier et al. 2008).

**4. Discussion**

What a scale analysis, like that by Liu et al. 2008, cannot do is predict or explain the dynamo mechanism that maintains the magnetic field in giant planets. Liu et al. consider only the axisymmetric *omega effect*, i.e., the shearing of their prescribed poloidal field into toroidal field by their prescribed "constant on cylinders" angular velocity in the outer region of the planet. They *assume* there is a 3D *alpha effect* somewhere deeper that twists toroidal field to maintain their prescribed poloidal field. They neglect the *alpha effect* in the outer region (where convection is presumably more vigorous) because they assume it is small there compared to the *omega effect*. However, as seen in 3D dynamo simulations, when integrated over a long time this persistent nonlinear feedback can modify the original poloidal field.

The main dynamo in a gas giant is likely acting at depths shallow enough for the shear flow and helical convection to be sufficiently vigorous and yet deep enough for the electrical conductivity to be sufficiently high. The axisymmetric parts of the flow and field could be close to a zero-ohmic-heating flow-field configuration mentioned above with most ohmic dissipation occurring in 3D at very small scales. Local ohmic heating drives small-scale 3D buoyant flow, which advects internal energy, momentum and field and so modifies the field and flow. Investigating all these subtle 3D nonlinear feedbacks (while producing less total ohmic heating than Jupiter's observed luminosity) is challenging enough using high-resolution 3D dynamically consistent MHD computer simulations, which satisfy conservation of mass, momentum and energy and are run long enough to allow the differential rotation and generated magnetic field to adjust. However, trying to accurately predict the dynamic nonlinear correlations and feedbacks of the unobserved 3D flow and field well below the surface using the steady-state axisymmetric scale-analysis approach Liu et al. have chosen is, as they admit, "entering uncertain territory."

The constraint on total ohmic heating is however an extremely useful test for 3D simulations of giant planets. Such models would require a realistic mass, radius, rotation rate and realistic radial stratifications of the mean density, temperature (e.g. Guillot 2005) and electrical conductivity (e.g. Liu et al. 2008). Because of computational limitations, the model's viscous and thermal diffusivities would likely need to have enhanced (turbulent) values, which requires a greater than realistic luminosity to drive fluid velocities and magnetic fields that are qualitatively similar to those observed on the planetary surfaces. The total ohmic heating in these 3D simulations (not only that from the axisymmetric current) can easily be calculated and monitored without any kinematic scaling arguments or assumptions about the structures and amplitudes of the flow and field configurations. The simulations do have many uncertainties; but the ohmic heating due to the simulated field and flow can be calculated exactly and compared to Jupiter's luminosity. Even if the model's luminosity needed to be much larger than the observed value, its ohmic heating compared to the observed luminosity provides a necessary constraint on the internal field and flow. If a simulation maintains a surface field and zonal wind pattern similar to that observed on Jupiter's surface (for example) but its ohmic heating exceeds Jupiter's luminosity the simulated subsurface field and flow would fail the test. All 3D computer simulations of the internal dynamics of giant planets should be tested this way.

These comments are *not* meant to suggest that the zonal winds observed on Jupiter or Saturn do extend into their deep interiors; the winds may indeed be only a shallow feature. However, the kinematic approach employed by Liu et al. 2008 produces results that are at best suggestive. It is not clear upon what they base their claims that their constraints on the zonal wind depths only have "uncertainties of a few percent" and "are insensitive to difficulties in modeling turbulent convection." Such constraints would require arguments that prove that there is no plausible configuration of field and flow that would allow strong zonal flow below such depths. Liu et al. do not provide such proof. Instead they provide estimates of the depth of zonal winds that would not violate the ohmic heating constraint for the configurations of field and flow that they choose, which are not derived from dynamically consistent MHD equations. As described above, an angular velocity nearly "constant on poloidal field lines" is a plausible, albeit still kinematic, counter example that would produce little ohmic heating and therefore, by the approach of Liu et al., would have no depth constraint. However, the ohmic heating analysis of Liu et al. does focus attention on this issue and suggests a challenging test for dynamically consistent 3D MHD modeling studies.

**Acknowledgments.** Support for this study was provided by grants from the NASA *Outer Planets Research Program* (NNG05GG69G) and the NASA *Solar and Heliospheric Physics Program* (NNG06GD44G).